# C-ITS Deployment in Europe - Current Status and Outlook

Katrin Sjöberg, Volvo

Peter Andres, Opel

Teodor Buburuzan, VW

Achim Brakemeier, Daimler

*Cooperative Intelligent Transport Systems (C-ITS) refer to applications using vehicle-to-vehicle and vehicle-to-infrastructure communications at a carrier frequency of 5.9 GHz to increase road traffic safety and road traffic efficiency in Europe (a.k.a. connected vehicle technology in the US). This article will shed some light on the current status of C-ITS in Europe and what is left before deployment can commence in 2019 as announced by C2C-CC. Even though there is an immense activity for the launch of C-ITS in Europe, the automotive industry is also planning for the future.*

## Introduction

Vehicle-to-vehicle (V2V) and vehicle-to-infrastructure (V2I) communications, collectively known as V2X communication, have great potential to increase traffic safety and efficiency. V2X communication is a wireless sensor that closes the gap between line-of-sight (LOS) sensors such as radars and cameras, and long-range cellular communication. LOS sensors cannot see beyond physical barriers and they cannot predict a traced object's intentions. The V2X sensor can overcome these shortcomings by providing status information about vehicles hidden around corners and behind other vehicles within milliseconds. Further, the V2X sensor can also receive information about the intention of other objects and the ego vehicle can adapt its behaviour. The V2X sensor beats the long-range cellular communication when it comes to local information dissemination in the immediate vicinity of the vehicles in milliseconds and the V2X sensor is not dependent on any base station or access point to function, i.e., no coverage is required from communication infrastructure and no subscriptions to a network operator is necessary.

In 2008, Europe received a frequency allocation at 5.9 GHz with the goal to increase road traffic safety and efficiency. This frequency allocation sparked the V2X activities in Europe. The available frequency band has been divided into 10 MHz communication channels, where the control channel (CCH) between 5.895-5.995 GHz will carry data traffic for increasing safety.

To achieve communication interoperability between implementations of different manufacturers, standardization plays an important role. In Europe, ETSI Technical Committee on Intelligent Transport Systems (TC ITS) has developed standards to support C-ITS day one applications. ETSI TC ITS has focused on protocols supporting applications on the vehicle side. Protocols supporting applications executed on smart infrastructure such as traffic lights, have been developed by CEN Technical Committee 278 working group 16 (TC 278 WG16). Smart infrastructure is using the same lower layer protocols as the vehicles.

Standardization forms the basis for deployment but not everything can be solved through standardization and protocol standards need to be parameterized and filled with relevant content. CAR 2 CAR Communication Consortium (C2C-CC) collects OEMs, suppliers, universities and research institutes, in Europe. C2C-CC plays an important role as an umbrella organization where OEMs and their partners can discuss topics related to V2V communication. V2V will only be leveraged if an interoperable system working across all OEMs and smart infrastructure is established. C2C-CC has created a basic system profile (BSP) to achieve interoperability based on the standardized protocols, and where gaps have been identified white papers have been compiled. C-ITS deployment in Europe is market-driven and C2C-CC has set the deployment start to 2019. BSP's counterpart in the US would be the recently published SAE J2945/1 addressing the minimum performance requirements for the V2V system.

## C-ITS protocol stack

The protocol stack for supporting road traffic safety applications using V2V communication is outlined in Figure 1. The European communications architecture is described in ETSI EN 302 665 and the protocol stack contains three layers – access-, networking & transport-, and facilities layer. The access layer merges the physical and the data link layer in the OSI model and the access layer technology is outlined in ETSI EN 302 663. It constitutes three parts: IEEE 802.11p, IEEE 802.2 logical link control (LLC), and ETSI TS 102 687 decentralized congestion control (DCC). The amendment IEEE 802.11p was enrolled in the compilation of a new version of IEEE 802.11 in 2012, but will hereafter be referred to IEEE

802.11p for easiness. Congestion control (DCC) is in place to restrict the number of packet transmissions when the network load increases to avoid unstable network behaviour.

In the networking & transport layer, GeoNetworking (ETSI EN 302 636-4-1) is used as a network protocol providing both singlehop as well as multihop communication through geographical addressing. In the multihop mode, GeoNetworking is facilitating routing based on geographical areas, e.g., a certain stretch of a road can be addressed. The basic transfer protocol (BTP) is a connectionless, best effort transport layer protocol providing means for distinguishing between different facilities layer protocols. It is outlined in ETSI EN 302 636-5-1. BTP is best effort since the data transmitted on behalf of traffic safety applications is of delay-sensitive character.

Cooperative awareness message (CAM) and decentralized environmental notification messages (DENM) are the two facilities layer protocols that have been standardized in ETSI EN 302 637-2 and ETSI EN 302 637-3, respectively. More information about CAMs and DENMs will be provided in subsequent sections. More details about the European protocol stack are found in [1]. The ETSI standards are available online for free.

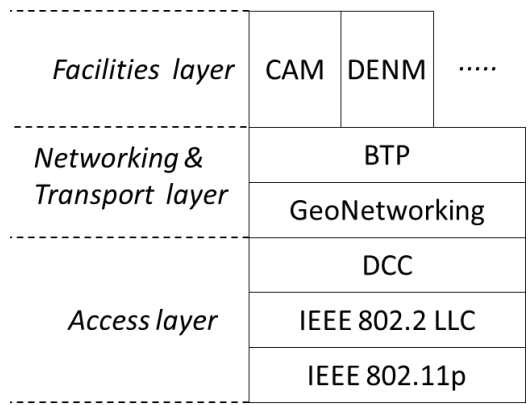

**Figure 1. C-ITS protocol stack.**

It should be noted that the C-ITS protocol stack follows a similar structure that is today found for communication via WiFi or Ethernet. LLC is providing means to distinguish between different network protocols via EtherTypes and GeoNetworking has its own EtherType. Due to this also the internet protocol (IP) is supported in the C-ITS protocol stack. Focus up until now have been to develop protocols for supporting road traffic safety (e.g., CAM and DENM) and BTP uses port numbers to identify different facilities layer protocols. In other words, by using the well-established ubiquitous WiFi standard IEEE 802.11 tailored to the vehicular environment a plethora of different applications can be supported (compare what your own laptop supports in terms of networking – emailing, video streaming, file transfer, web browsing, IP telephony). The C-ITS protocol stack is not a silo implementation and it is just the imagination that put up limits.

**Day one applications**

The artery of C-ITS is the position messages (a.k.a. "Here I am", beacons) broadcasting information about the ego vehicle such as speed, position, and heading, etc. The official protocol name for this position message in Europe is cooperative awareness message (CAM) and in the US, this is called basic safety message (BSM). The CAMs will be transmitted with 1-10 Hz depending on vehicle dynamics (see ETSI EN 302 637-2) and those will always be transmitted to increase awareness among traffic participants. Decentralized environmental notification messages (DENMs) are event-triggered messages broadcasted in case of a noteworthy event. As long as the event is valid, DENMs will be broadcasted alongside with CAMs. These two facilities layer protocols support day one applications for vehicles. Examples of day one applications triggered by vehicles are stationary vehicle warning, slow vehicle warning, emergency electronic brake light, emergency vehicle approaching, adverse weather conditions, etc. In C2C-CC, triggering conditions for a set of day one applications have been developed and those have not yet been put into standardization. The triggering of CAMs is described in EN 302 637-2.

CAMs and DENMs do not contain any vehicle identification number (VIN) or any data about the driver or the brand of the vehicle. CAMs contain what kind of vehicle type that is broadcasting the information (e.g., car, bus, truck, etc.). DENMs contain information about the noteworthy event itself and its attributes such as position, speed, and heading, if applicable. DENMs are comparable to event flags in the BSM.

The intention of day one applications is to increase the information horizon for the driver, in other words, they are a driver support function. Standardization has focused on the transmitting side and left the receiving side to be implementation specific to the extent possible. Nothing is preventing an OEM to use data received from other vehicles to control the vehicle but this is under the responsibility of every OEM. Competitiveness between brands is achieved on the receiving side.

*Preparation for deployment*

To prepare for deployment, a BSP has been compiled by the members of C2C-CC to create an interoperable system that can function across brands. However, the BSP together with the standardized protocols and white papers do not solve all practical issues to facilitate a deployment start. The V2V security framework requires a public key infrastructure (PKI) system to be established providing the root public key for all European vehicles. In the V2V system, all vehicles will have a public key and a set of short-term private keys. A short-term private key (pseudonym) is used to sign an outgoing message and the public key is used to verify an incoming message with.

If the verification is successful, then the data has not been modified since the signing (data integrity is preserved) and the sender can be trusted (data origination). However, the signing of outgoing messages does not reveal if the data put into the message is correct or not and therefore, plausibility checks must also be incorporated on the sending and receiving sides, respectively. The private key (pseudonym) will only be valid for a certain period of time to preserve the privacy of the vehicle and then it is changed. ETSI TC ITS is currently investigating different pseudonym change strategies. More details about the security framework can be found in [7].

To be part of the security framework, the requirements set out in the BSP must be fulfilled through conformance testing. Once an OEM has passed the compliance assessment, access to the PKI is granted. Compliance assessment and the security framework go hand in hand and both need to be resolved before deployment can commence. For initial deployment, self-certification has been proposed to minimize costs.

However, to put radio equipment on the European market, the harmonized EN 302 571 has to be fulfilled. This standard developed by ETSI, outlines requirements on the radio transceiver such as output power level, spectrum mask, etc., to avoid disturbing already existing services in neighbouring frequency bands. Further, it also puts up requirements on co-existence between electronic toll collection using a frequency band at 5.8 GHz and C-ITS using 5.9 GHz (through normatively referencing ETSI TS 102 792), and decentralized congestion control.

## 5.8 GHz and 5.9 GHz co-existence

Electronic toll collection (ETC) in Europe is using a frequency band at 5.8 GHz together with standards developed within CEN, and it goes under the epithet CEN DSRC (dedicated short-range communication). The notion of DSRC has been used for ETC in Europe for over two decades but to avoid confusion with the US DSRC referring to IEEE 802.11p, CEN was added in front of DSRC. CEN DSRC is a simple radio frequency identification (RFID) system, where a roadside unit (RSU) emits 2W to wake-up the on-board unit (OBU) when the vehicle is passing by for collecting the fee. The OBU answers by using the energy coming from the RSU (backscatter technology). The OBU is a simple transceiver with no blocking capability – limited possibility to discriminate between wanted and unwanted signals. The non-existing blocking capability makes the CEN DSRC OBU sensitive to the communication at 5.9 GHz. It has been identified when the number of C-ITS equipped vehicles increases that the sensitive toll transaction might fail.

The requirements set out in EN 302 571 are legal requirements and therefore, co-existence methods have been developed and are outlined in ETSI TS 102 792. First, the vehicle must know when it is in a tolling zone, and this can be performed using one of the two following methods; (*i*) database containing all toll plazas throughout Europe, or (*ii*) CEN DSRC radio detector. A vehicle must also react to a special message broadcasted on 5.9 GHz signalling the presence of an upcoming tolling zone (this is an optional feature that ETC operators can provide for new installations of tolling zones). Once in a tolling zone, the vehicle needs to adapt its output power and duty cycle depending on the number of other transmitting vehicles within radio range and the transceiver's specific radio characteristics.

## Wireless Performance

The wireless performance is of crucial importance to reach the goals of increased road traffic safety and efficiency. Electromagnetic wave propagation at a carrier frequency of 5.9 GHz implies a high multipath environment (several replicas of the signal are reaching the receiver due to the bouncing on objects in the environment) and the signal can easily be blocked. Also the high relative speeds that can be reached in highway scenarios can be challenging.

C2C-CC is currently working on adding wireless performance requirements to the BSP. One general objective that has been proposed is that the vehicle shall be able to receive 90% of all packets at a distance of 400 meters omnidirectionally in an interference-free environment, given: (*i*) a packet length of 400 bytes, (*ii*) a transfer rate of 6 Mbps, and (*iii*) an output power of 23 dBm (e.i.r.p).

For testing performance on radio systems, channel models are used. Channel models try to emulate what the real world is exposing the signals to. Channel models are always a trade-off between complexity and capturing the essence of the real-world. When radio hardware is under development, those are benchmarked towards standardized channel models. However, channel models for V2V communication are more difficult than for traditional radio systems because both transmitter and receiver are moving (can have relative high speed differences) and antennas are approximately at the same height. C2C-CC has specified five channel models representing 5 different scenarios related to LOS and non-LOS conditions in urban, rural, and highway environments. These models can be used for testing the radio on a component level together with a single antenna. They are not suitable for a hardware-in-the-loop (HIL) testing. Therefore, improved channel models are currently under investigation in C2C-CC that can be used for testing the whole system and not just the characteristics of the radio hardware, but also the impact of a multi-antenna system with transmit and receive diversity.

When C-ITS deployment commence, one frequency channel will be used for disseminating CAM and DENM. But once deployment starts, the uptake to start using neighbouring frequency channels for day two and day three applications will be quick. Then the adjacent and alternate channel rejection together with the blocking capabilities for the transceiver will come into play. All these three capabilities are revealing to what extent a receiver can discriminate between the wanted signals on a specific frequency channel tuned into in the presence of on-going transmissions on neighbouring frequency channels. The requirements on the adjacent and alternate adjacent channel rejection are outlined in EN 302 571, but those values stem from the IEEE 802.11 standard used for WiFi operation. However, there is a need to revise these values in the next version of EN 302 571 to increase the performance of C-ITS when using several frequency channels.

### AUTOSAR

AUTomotive Open System ARchitecture (AUTOSAR) is a partnership between OEMs and suppliers. In short, AUTOSAR provides standards for basic software functionality of automotive ECUs (similar to middleware software) to facilitate scalability, traceability, to support different functional domains, and to establish an open architecture. Support for European V2X has been introduced to the Classic Platform of AUTOSAR in its latest Release 4.3 [9].

### Change control board

As mentioned earlier, C2C-CC has released an initial version of the BSP with the same intention as SAE J2945/1, namely, to create an interoperable system out of standards for deployment. The BSP is a common effort among the different working groups within C2C-CC, where representatives from OEMs, suppliers, universities, and research institutes are contributing to the work.

With the release of the first version of BSP, containing the requirements, it is a necessity to synchronise the work of the different working groups and keep track of change proposals that needs to be included in future releases of the BSP. In other words, a strong need for traceability and version control of the requirements has been identified. Thus, C2C-CC has established a change management system (CMS), where every member can send in and document changes for existing requirements or new requirements in the BSP. These proposals are then assigned to the responsible working group for expert discussion. The discussion is documented in the CMS as well as the final conclusion or proposal. The final agreement on the change is then done by a Change Control Board (CCB), which accepts or rejects the change. The CCB also decides about the new release cadence of the BSP. Versioning control and change management are an integral part of standardization but this is also a necessity for deployment of systems.

While the C2C-CC will start deployment in 2019, it is not expected that all partners will launch at the same time. More likely is a deployment across several years. Therefore, the CMS is not only important to keep track of changes, but is critical to ensure interoperability between different releases of the BSP. Only then it is possible that all vehicles/smart infrastructures are interoperable. But it also guarantees a way to continue to develop and enhance the technology.

### Day Two and Beyond

With the advent of automated driving functions, especially with the broad availability of vehicles capable of supporting higher automation levels (3-5), the need for cooperation and coordination between the various traffic participants becomes increasingly necessary. All automated vehicles rely on the premise that they continuously plan their trajectories and, based on the observed environment, select one or another as the current driving trajectory. Currently, this requires a major overhead for unpredictable behavior since it is not 100% certain what another vehicle, or another traffic participant, will do in the next several seconds. That is why relatively large "buffers" have to be included in these trajectories, especially when planning them around other moving vehicles. If these other vehicles would share, or even constantly disseminate their own plans, other vehicles could use this information to reduce the uncertainties and so minimize the buffers within their trajectories. This would enable automated driving vehicles to drive closer to each other (and so increase the capacity of roads and cities), react more quickly to maneuvers, be better controlled, and avoid collisions.

Based on the initial deployment of the IEEE 802.11p technology, the members of the C2C-CC, representing the large part of the automobile industry, have created a staged deployment strategy using a development roadmap structuring the past, current and future research and standardization work in the field of communicating and cooperative vehicles.

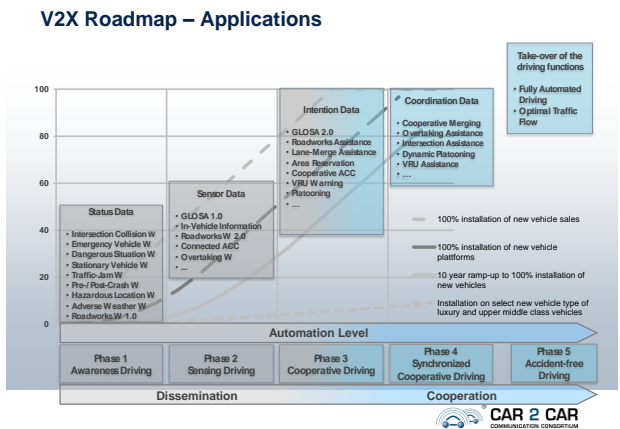

Figure 2. C2C-CC Applications Roadmap

The C2C-CC applications roadmap, presented in Figure 2, envisions four deployment phases for direct V2V communication. Each subsequent phase extends the previous one by allowing vehicles to exchange more information, thus enabling new applications and classes of use cases to be realized. Each new phase is characterized by the new type of information it allows traffic participants to disseminate and share between themselves:

1. The first, initial phase, as described in the previous chapters, will enable vehicles to

disseminate their *status* information, thus allowing other vehicles to become aware of their presence and of eventual hazards detected on the road.
2. The second, sensing driving phase will allow various traffic participants to provide additional information, namely information gained from the various on-board *sensors* such as cameras and radars. This additional information enables vehicles to "see with the eyes of others" and so detect otherwise hidden objects (e.g., around a corner), or get a more accurate view on what is happening within their environment (e.g., an intersection with various vehicles and pedestrians).
3. The third, cooperative driving phase will allow vehicles to share their *intentions* with other traffic participants, and so provide them with a glimpse into the individual future of each vehicle. Information such as trajectories or planned maneuvers will be used by automated driving algorithms to enable vehicles to accurately predict what other traffic participants will do in the near future and so optimize their own decisions and maneuvers.
4. The last, *synchronized* driving phase is where vehicles are autonomously driven through almost all situations (levels 4 and 5 in Figure 2) and are able to exchange and synchronize driving trajectories among each other and so achieve optimal driving patterns.

One question that can be raised when looking at the above roadmap is whether there are special requirements on vehicle communication for higher automation levels. For example, if a collision has to be prevented due to an unexpected event, the vehicles have to act autonomously just before the collision takes place. The vehicles would not only exchange trajectories, but also continuously re-adjust them (cooperative decision making). Sensor data exchange for collective perception will lead to larger message sizes. New messages carrying intention and coordination data for cooperative agreement will need to be defined. Finally, the introduction of high automation based on V2X connectivity demands appropriate security requirements to be addressed as well.

## CACC and platooning

Cooperative adaptive cruise control (CACC) and platooning are two promising applications for the future. Both are dependent on the use of IEEE 802.11p technology. In other words, the V2V communication is the enabler for pulling the vehicles together and reduces the distance between the vehicles. There are two main advantages with reducing the distance and a third is an outcome of the second: (*i*) more vehicles can be squeezed into the existing road network, and (*ii*) a reduction in fuel consumption can be achieved due to reduced air drag. The major outcome of the latter is reduced environmental impact. This is especially pronounced for heavy duty vehicles.

In CACC, the vehicle is only controlled longitudinally and the driver has its hands on the steering wheel. It is an enhanced ACC where the ego vehicle receives information wirelessly from vehicles upstream and information about changes in acceleration can be received before the in-vehicle radar can detect it. Research on CACC has shown that the capacity on existing road network can almost be doubled if all vehicles were V2V equipped [2, 3].

In platooning, also known as road trains, vehicles are controlled both laterally as well as longitudinally. Platooning is an interesting application especially for trucks to reduce fuel consumption. The first truck in the platoon is driven manually and the rest of the trucks will be automatically controlled based on data received wirelessly from the leading truck together with on-board sensors. Up to 20% of fuel can be saved in platoons in best case [4] but fuel savings are highly dependent on spacing between vehicles, load, road topology, and geometry of vehicles.

ETSI TC ITS has initiated standardization work on CACC and platooning. Two technical reports (TR) are under development with a planned release date of March 2017. The aim with the two TRs is to identify standardization needs from a communication perspective.

Platooning was successfully showcased in Europe in the beginning of April 2016. It was the government of the Netherlands, currently holding the presidency of EU, which had invited all truck OEMs to Rotterdam for a large-scale demonstration of platooning called "European Truck Platooning Challenge" [5]. The Netherlands initiated this event to shed some light on future intelligent transport systems.

## Functional Safety

Day one applications are increasing the information horizon for the driver and the vehicle will not be automatically controlled using data received on the wireless channel. This means, it is a driver support function informing the driver about situations and is not subject to a rigorous functional safety assessment. When moving to day two applications and beyond, information received from the V2X sensor will be used for direct control of the vehicle (e.g., CACC and platooning). This will put up new requirements on the functional safety.

The standard ISO 26262 is addressing functional safety of internal electrical and electronic systems of vehicles. It is a working methodology including amongst other things a hazard analysis, where a hazardous event is classified according to severity, exposure, and controllability. ISO 26262 provides a way of breaking down the complexity of the electrical architecture of a vehicle and means to achieve safety goals.

However, applications such as CACC and platooning are not only concerning the internal electrical architecture of the vehicle but also information is received from other vehicles that shall be used by the internal architecture to control the vehicle. CACC and platooning are classified as a system-of-systems due to: (*i*) operational independence (the vehicle can operate outside the application), (*ii*) managerial independence (each vehicle could have a different owner and developer), (*iii*) evolutionary independence (the development of each vehicle is not synchronized), and (*iv*) emergent behaviour (no vehicle can achieve the goal of the application in question by itself, e.g., traffic efficiency and fuel savings). The functional safety of system-of-systems for road vehicles has not been studied thoroughly. There exist different functional safety approaches stemming from other industries but those do not perfectly match the needs of road vehicles. Since ISO 26262 is the standard for the internal architecture, it would be beneficial to extend the ISO 26262 framework to address also system-of-systems.

## European Outlook

The European Commission (EC) has recognized that C-ITS has the potential to increase road traffic safety and road traffic efficiency. The EC has funded research for over a decade on different aspects of C-ITS and mobility through what is called Framework Programmes (FPs). In FP7, run between 2007 and 2013, some large C-ITS research projects were executed that laid the foundation for standardization and interoperability. Currently, policy recommendations for C-ITS deployment in Europe are under development and this work was initiated by the EC. The work is called "C-ITS deployment platform" and several departments of the EC are part of it together with stakeholders from national authorities and industry (e.g., OEMs, suppliers, telecom operators). The overall goal of the platform is to provide policy recommendations to the EC by identifying main barriers for C-ITS deployment. The first phase of this work was ended in January 2016 with a report summarizing the stakeholders' contributions [6].

Generally, C-ITS deployment is suffering from the chicken and egg problem. OEMs need to invest for at least 5-10 years prior to customers can see the benefit of all C-ITS applications. Even if all new vehicles were equipped, it will take some years before a reasonable market penetration is reached. Certain C-ITS applications need higher penetration rates than others to function properly. Customers would see an immediate benefit if applications run on smart infrastructure such as variable messages signs and road works warning were in place. There is a major project in Europe since a couple of years called the European C-ITS corridor. It is a smart-road deployment project involving road authorities in The Netherlands, Germany, and Austria [8]. The goal with this common effort is to have smart infrastructure informing drivers about road works and other obstacles on the road between Amsterdam via Frankfurt to Vienna. Penetration rates would not be as critical, if smart infrastructure would be in place supporting drivers.

Applications such as CACC and platooning are more dedicated towards traffic efficiency and goods transports (heavy duty vehicles) and they might find its way quicker into the market than pure safety related features. These applications are niche markets, which are not dependent on a high penetration rate.

Lately, C-ITS deployment has also been challenged by other unforeseen and unpredictable sources of concern. The WiFi industry wants to share the allocated frequency band at 5.9 GHz both in Europe and in the US with vehicles and smart infrastructure. Sharing is only possible if the V2X communication is not affected at all. The C-ITS deployment is further blurred by all efforts and money currently put into the 5G development. The EC has set aside 700 MEUR through its FP Horizon 2020 to research, development and innovation of 5G technologies [10].

## Conclusions

In the past safety has focussed on helping people to survive a crash, C-ITS is about avoiding crashes and increasing traffic efficiency. The wireless sensor IEEE 802.11p can "see" beyond physical barriers and provide information about noteworthy events to the driver in the immediate vicinity of the vehicle within milliseconds. It is a technology closing the gap between LOS sensors such as cameras and radars and long-range cellular technology. IEEE 802.11p uses a dedicated frequency band at 5.9 GHz, which is royalty-free implying that no expensive subscription to a network operator is necessary.

C2C-CC has developed a basic system profile (BSP) by parameterizing the C-ITS standards that have been developed within ETSI TC ITS. The BSP facilitates an interoperable system. Missing parts before C-ITS deployment can commence are the setup of the security framework (PKI) that holds the public key and nitty-gritty details on how certification shall be performed for a vehicle installation. In short, there will be three steps for an OEM to put a C-ITS equipped vehicle onto the European market. First, the mandatory radio standard ETSI EN 302 571 must be fulfilled. Secondly, the vehicle needs to pass the C2C-CC certification, i.e., the testing of the requirements set out in the BSP. If the C2C-CC certification is passed, the V2V-equipped vehicle will gain access to the PKI and will be a trusted vehicle, fulfilling the minimum performance requirements.

When moving to higher automation levels, V2V communication facilitates the exchange of intentions and high-resolution sensor data between vehicles in a millisecond time frame. This will allow the IEEE 802.11p technology to move from being a wireless sensor into a wireless actor, which will be able to not only sense its surrounding but also interact and change it. The exchange of data will make the journey smoother for the automated vehicle since the control of the vehicle in relation to other vehicles will be more precise, predicted and synchronized.

To develop a new wireless system from farm-to-table takes around 10 years including research, field-operational tests, and standardization. In Europe, standardization started in 2008 following the frequency allocation; this suggests that commercial deployment could be initiated in 2018. The 10-year cycle is also true for the development of new generations of cellular technology.

C-ITS deployment is around the corner and many activities are on-going in Europe. Penetration rate of V2V-equipped vehicles for certain applications are crucial but smart infrastructure applications and niche applications such as CACC and platooning are not dependent on high penetration rates. Unfortunately, OEMs need currently to spend many resources on defending their access to the allocated frequency band at 5.9 GHz and the selected wireless technology (IEEE 802.11p); those resources could come much better off focussing on deployment issues.

*Dr. Katrin Sjöberg works as a connected vehicle technology specialist at Volvo Group Trucks Technology in Göteborg, Sweden. Her research interests range from channel modelling to applications for wireless systems in general and for V2X communication in particular. She is active in C2C-CC and in ETSI TC ITS, where she holds a vice chairmanship of WG4.*

*Dipl-Ing Peter Andres is a System Engineer for V2X communication at GM/Opel located in Ruesselheim, Germany. His focus is on the global deployment of vehicle-to-vehicle communication technology. He is also the chair of the WG Deployment of the C2C-CC.*

*Dr. Teodor Buburuzan is a senior researcher for the Volkswagen Group Research in Wolfsburg, Germany. His research interests are centred around connected and automated driving topics with clear focus on V2X communication for cooperative systems. He currently chairs C2C-CC WG Roadmap and ETSI TC ITS WG1.*

*Dr. Achim Brakemeier is a manager for Car2x vehicle based functions at Daimler AG. He is active in the field of driver assistance systems with a special focus on wireless systems. He currently chairs the WG Communication in C2C-CC.*